\documentclass[letterpaper, 10pt, conference]{ieeeconf}
\IEEEoverridecommandlockouts
\usepackage{amsmath, amssymb}
\usepackage{authblk}

\usepackage[colorlinks=true]{hyperref}
\newtheorem{lemma}{Lemma}
\newtheorem{remark}{Remark}

\newcommand{\ba}{\boldsymbol{a}}

\newcommand{\bD}{\boldsymbol{D}}

\newcommand{\bK}{\boldsymbol{K}}
\newcommand{\bL}{\boldsymbol{L}}

\newcommand{\bP}{\boldsymbol{P}}

\newcommand{\bxi}{\boldsymbol{\xi}}

\newcommand{\bSz}{\boldsymbol{\sigma_{\! z}}}

\newcommand{\cD}{\mathcal{D}}
\newcommand{\cM}{\mathcal{M}}
\newcommand{\CC}{\mathbb{C}}

\newcommand{\ket}[1]{\left|#1\right\rangle}

\newcommand{\braket}[2]{\left\langle #1\right|\left.#2 \right\rangle}
\newcommand{\ketbra}[2]{\left|#1 \right\rangle\!\left\langle #2 \right|}

\newcommand{\Tr}[1]{\mathrm{Tr}\!\left(#1\right)}

\newcommand{\EE}[1]{\mathbb{E}\!\left[#1\right]}

\title{\LARGE \bf Diffusive Stochastic Master Equation
	\\
	with Dispersive Qubit/Cavity Coupling%
	\thanks{This project has received funding from the European Research
		Council (ERC) under the European Union's Horizon~2020 research and
		innovation programme (grant agreement No.~884762).}}

\author{Pierre Rouchon%
\thanks{P.~Rouchon is with Laboratoire de Physique de l'\'Ecole Normale
        Sup\'erieure, Mines Paris--PSL, Inria, ENS-PSL, Universit\'e PSL,
        CNRS, Paris, France.
        {\tt\small pierre.rouchon@minesparis.psl.eu}}}

\begin{document}

\maketitle
\thispagestyle{empty}
\pagestyle{empty}

%%%%%%%%%%%%%%%%%%%%%%%%%%%%%%%%%%%%%%%%%%%%%%%%%%%%%%%%%%%%%%%%%%%%%%%%%%%%%%%%
\begin{abstract}
We provide a detailed analysis of the quantum diffusive Stochastic Master
Equation~(SME) for qubit/cavity systems with dispersive coupling, incorporating
classical input signals and output signals (measurement outcomes via
homodyne/heterodyne detection).  We show that the dynamics of the
qubit/cavity density operator converges exponentially towards a slow invariant
manifold.  This manifold is parameterized by the density operator of a
\emph{fictitious} qubit governed by a quantum SME that incorporates the
classical input/output signals while preserving complete positivity and trace.
The reduced cavity (resp.\ qubit) operator, obtained by partial trace over
the qubit (resp.\ cavity), is then expressed as a time-varying deterministic
quantum channel acting on the fictitious-qubit density operator.  This
formulation avoids the non-Markovian descriptions---with temporarily negative
dephasing rates or detection efficiencies exceeding one---that appear in
several publications.  Extensions are considered in which the qubit is replaced
by a qudit dispersively coupled to an arbitrary set of cavity modes driven by
collective classical input/output signals.
\end{abstract}

\section{Introduction}

Quantum non-demolition (QND) measurements~\cite{wiseman-milburnBook,%
haroche-raimondBook06} play a central role in quantum error correction.
Syndrome detection is based on joint parity measurements~\cite{nielsen-chang-book}.
For superconducting qubits these are implemented with homodyne/heterodyne setups
(see, e.g.,~\cite{Govia2015ScalableTA,BlaisRMP2021}), whose dynamics are governed
by quantum Stochastic Master Equations (SMEs; for a mathematically rigorous
treatment see~\cite{BarchielliGregorattiBook}; for books covering continuous
measurement and feedback see~\cite{wiseman-milburnBook,JacobsBook2014}; and for tutorial
introductions see~\cite{Gough2018,ARC_PR_2022}).
Landmark experiments have demonstrated the real-time tracking of individual quantum
trajectories of superconducting qubits under dispersive homodyne
measurement~\cite{MurchNature2013,VijayNature2012}.

These models share a common structure first explored in~\cite{GambeBBHSG2008PRA},
where a \emph{polaron transformation} (a time-varying unitary change of frame) is
employed.  There, the ensemble-average evolution of the qubit density operator is
shown to obey a Lindblad-type master equation in which the dephasing rate can be
temporarily negative (the term $\gamma_\phi+\Gamma_d(t)$ in
\cite[eq.~(3.10)]{GambeBBHSG2008PRA}), a signature of non-Markovian dynamics in
the physics literature.  Nonlinear corrections to the dispersive regime, and their
impact on qubit dephasing and relaxation, are analysed in~\cite{BoissonneaultPRA2009}.
Several subsequent works follow this polaron approach with different
setups~\cite{Korotkov2016QuantumBA,CrigerEPJQT2016,SteinmetzSiddiqiJordanPRA2022}.
In particular, extensions to qudits with several cavity modes have been studied using
generalized polaron transformations, but the non-Markovian character of the qudit
dynamics persists (see, e.g., \cite[eqs.~(47--49)]{Yu2023StochasticMO}).

We propose here an alternative formulation for the evolution of the qudit density
operator.  It is based on a fully Markovian evolution---either deterministic or
stochastic---of a \emph{fictitious} qudit, from which the physical qudit density
operator is recovered via a time-varying quantum channel
(see~\eqref{eq:qubitoperator} for the qubit/cavity case).  This formulation:
\begin{itemize}
  \item avoids negative dephasing rates and detection efficiencies exceeding one;
  \item relies on an invariant manifold (Lemma~\ref{lem:qubit}) that is
        exponentially attractive at a rate set by the measurement strength
        (Lemma~\ref{lem:qubitconvergence});
  \item provides an exact reduced-order model that preserves complete positivity.
\end{itemize}
From a control-theoretic viewpoint, the quantum channel in~\eqref{eq:qubitoperator}
acts as an \emph{output map}: $\rho_Q$ (the physical qubit state, i.e., the output)
is obtained from $\bxi_Q$ (the fictitious qubit state, i.e., the internal state
governed by the Markovian SME~\eqref{eq:dynxiQ}).  This connection between quantum
stochastic dynamics and standard state-space representations opens new avenues for
quantum control design, ranging from quantum error correction to precision metrology.

\medskip
\textbf{Outline.}  Section~\ref{sec:qubitcavity} treats the qubit/cavity system of
\cite{GambeBBHSG2008PRA} in which the dephasing rate can be negative.
Section~\ref{sec:qudit} addresses the qudit extension of~\cite{CrigerEPJQT2016}.
Section~\ref{sec:multimode} covers the multi-cavity qudit case
of~\cite{CrigerEPJQT2016,Yu2023StochasticMO}.  Throughout, we focus on homodyne
detection; analogous results hold for heterodyne detection.

\section{The qubit/cavity case} \label{sec:qubitcavity}

\subsection{Model and change of frame}

Following~\cite{GambeBBHSG2008PRA}, the SME for the density operator~$\rho$ of
the composite qubit/cavity system reads
\begin{multline}\label{eq:dynrho}
  d\rho = -i\!\left[\sqrt{\kappa}\,u\,\ba^\dag + \sqrt{\kappa}\,u^*\ba
            + \chi\,\bSz\,\ba^\dag\ba,\;\rho\right]dt \\
         + \kappa\,\cD_{\ba}(\rho)\,dt
         + \sqrt{\eta\kappa}\,\cM_{\ba}(\rho)\,dw,
\end{multline}
where the classical input signal $u(t)\in\CC$ drives the cavity coherently,
and the homodyne measurement outcome satisfies
\[
  dy_t = \sqrt{\eta\kappa}\,\Tr{\ba\rho+\rho\ba^\dag}\,dt + dw_t.
\]
Here $\ba$ is the photon annihilation operator, $\bSz$ the qubit Pauli-$z$
operator, and $dw$ a standard Wiener increment.  The superoperators are
defined by
\[
  \cD_{\bL}(\rho) \equiv \bL\rho\bL^\dag - \tfrac{1}{2}(\bL^\dag\bL\rho
  + \rho\bL^\dag\bL),
\]
\[
  \cM_{\bL}(\rho) \equiv \bL\rho + \rho\bL^\dag
  - \Tr{\bL\rho+\rho\bL^\dag}\rho.
\]
The real parameters are: $\chi$ (dispersive coupling strength),
$\kappa>0$ (cavity decay/measurement rate), and $\eta\in[0,1]$
(homodyne detection efficiency).

\medskip
As in~\cite{GambeBBHSG2008PRA}, we introduce the polaron change of frame
\begin{equation}\label{eq:changeframe}
  \rho = \bigl(\bP_g\bD_{\alpha_g}+\bP_e\bD_{\alpha_e}\bigr)\,
         \bxi\,
         \bigl(\bP_g\bD_{-\alpha_g}+\bP_e\bD_{-\alpha_e}\bigr),
\end{equation}
where $\bP_g=\ketbra{g}{g}$, $\bP_e=\ketbra{e}{e}$ are the ground- and
excited-state projectors, and
$\bD_\alpha=e^{\alpha\ba^\dag-\alpha^*\ba}$ is the displacement operator
of amplitude~$\alpha\in\CC$.  The time-varying amplitudes $\alpha_g(t)$,
$\alpha_e(t)$ follow the linear ODEs
\begin{align}
  \dot\alpha_g &= i\chi\alpha_g - i\sqrt{\kappa}\,u
                  - \tfrac{\kappa}{2}\alpha_g,
                  \label{eq:dynalphag}\\
  \dot\alpha_e &= -i\chi\alpha_e - i\sqrt{\kappa}\,u
                  - \tfrac{\kappa}{2}\alpha_e.
                  \label{eq:dynalphae}
\end{align}
Note the sign difference in front of $\chi$: the two coherent pointers rotate
in opposite directions in phase space, encoding the qubit state.  Substituting
into~\eqref{eq:dynrho} and using Ito calculus yields, after straightforward
but tedious algebra. Defining the shorthand $\bL \equiv \ba+\alpha_g\bP_g+\alpha_e\bP_e$,
{\scriptsize
\begin{multline}\label{eq:dynxi}
  d\bxi =
  -i\!\left[\bP_g\!\left(\Re(\alpha_g\sqrt{\kappa}\,u^*)
            - \chi\ba^\dag\ba
            + \tfrac{i\kappa}{2}(\alpha_g\ba^\dag-\alpha_g^*\ba)
            \right),\bxi\right]dt \\
  -i\!\left[\bP_e\!\left(\Re(\alpha_e\sqrt{\kappa}\,u^*)
            + \chi\ba^\dag\ba
            + \tfrac{i\kappa}{2}(\alpha_e\ba^\dag-\alpha_e^*\ba)
            \right),\bxi\right]dt \\
  + \kappa\,\cD_{\bL}(\bxi)\,dt
  + \sqrt{\eta\kappa}\,\cM_{\bL}(\bxi)\,dw,
\end{multline}
}
with measurement output
\[
  dy = \sqrt{\eta\kappa}\,\Tr{\bL\,\bxi + \mathrm{h.c.}}\,dt + dw.
\]

\subsection{Invariant manifold and reduced-order model}

The key observation is that the vacuum component of the cavity decouples exactly
on a certain invariant submanifold.

\begin{lemma}[Invariant manifold]\label{lem:qubit}
  If $\bxi(0)=\bxi_Q^0\otimes\ketbra{0}{0}$, then for all $t>0$,
  \[
    \bxi(t) = \bxi_Q(t)\otimes\ketbra{0}{0},
  \]
  where the \emph{fictitious qubit} state $\bxi_Q$ is governed by the
  standard (Markovian) SME
  \begin{multline}\label{eq:dynxiQ}
    d\bxi_Q =
    -i\!\left[\Re(\alpha_g\sqrt{\kappa}\,u^*)\bP_g
              +\Re(\alpha_e\sqrt{\kappa}\,u^*)\bP_e,\;\bxi_Q\right]dt \\
    + \kappa\,\cD_{\alpha_g\bP_g+\alpha_e\bP_e}(\bxi_Q)\,dt
    + \sqrt{\eta\kappa}\,\cM_{\alpha_g\bP_g+\alpha_e\bP_e}(\bxi_Q)\,dw,
  \end{multline}
  with output
  \[
    dy = \sqrt{\eta\kappa}\,
         \Tr{(\alpha_g\bP_g+\alpha_e\bP_e)\bxi_Q+\mathrm{h.c.}}\,dt + dw,
  \]
  and initial condition $\bxi_Q(0)=\bxi_Q^0$.
\end{lemma}

In other words, the set
$\mathcal{I}_Q = \{\bxi_Q\otimes\ketbra{0}{0}\mid \bxi_Q \text{ qubit density operator}\}$
is invariant under~\eqref{eq:dynxi} for every Wiener realization.

\begin{lemma}[Exponential convergence]\label{lem:qubitconvergence}
  Every solution of~\eqref{eq:dynxi} converges exponentially and almost
  surely to $\mathcal{I}_Q = \{\bxi\mid\Tr{\ba^\dag\ba\,\bxi}=0\}$.
  More precisely, the mean photon number satisfies
  \[
    \frac{d}{dt}\EE{\Tr{\ba^\dag\ba\,\bxi}} = -\kappa\,\EE{\Tr{\ba^\dag\ba\,\bxi}},
  \]
  so $\Tr{\ba^\dag\ba\,\bxi}$ is a super-martingale decaying at rate~$\kappa$.
\end{lemma}

\begin{remark}
The proof relies on formal stochastic calculus; a rigorous functional-analytic
justification---ensuring well-posedness and exponential convergence in appropriate
operator-norm topologies---can be obtained within the framework of~\cite{ChebotarevFagnola1998}.
\end{remark}

\subsection{State-space input/output representation}

We now assemble the preceding results into a complete input/output description.
Suppose the initial qubit/cavity state is
$\rho(0)=\rho_Q^0\otimes\ketbra{\alpha^0}{\alpha^0}$ for some
$\alpha^0\in\CC$.  The full system is then described by:
\begin{enumerate}
  \item the ODEs~\eqref{eq:dynalphag}--\eqref{eq:dynalphae} with initial
        condition $\alpha_g(0)=\alpha_e(0)=\alpha^0$;
  \item the SME~\eqref{eq:dynxiQ} with initial condition $\bxi_Q(0)=\rho_Q^0$.
\end{enumerate}
The full density operator at time~$t$ is recovered from the
state $(\alpha_g(t),\alpha_e(t),\bxi_Q(t))$ via the output
map~\eqref{eq:changeframe}:
\begin{multline*}
  \rho(t) = \bigl(\bP_g\bD_{\alpha_g(t)}+\bP_e\bD_{\alpha_e(t)}\bigr)
            \bigl(\bxi_Q(t)\otimes\ketbra{0}{0}\bigr) \\
            \times\bigl(\bP_g\bD_{-\alpha_g(t)}+\bP_e\bD_{-\alpha_e(t)}\bigr).
\end{multline*}
Tracing over the qubit yields the cavity state
\begin{multline*}
  \rho_C(t) = \Tr{\bP_g\bxi_Q(t)}\ketbra{\alpha_g(t)}{\alpha_g(t)} \\
             + \Tr{\bP_e\bxi_Q(t)}\ketbra{\alpha_e(t)}{\alpha_e(t)}.
\end{multline*}
Tracing over the cavity yields the qubit state
\begin{multline}\label{eq:qubitoperator}
  \rho_Q(t) = \bigl(\bP_g+c_{ge}(t)\bP_e\bigr)\bxi_Q(t)
              \bigl(\bP_g+c_{ge}^*(t)\bP_e\bigr) \\
             + \bigl(1-|c_{ge}(t)|^2\bigr)\bP_e\,\bxi_Q(t)\,\bP_e,
\end{multline}
where the overlap between the two coherent pointer states is
\[
  c_{ge}(t) = \braket{\alpha_g(t)}{\alpha_e(t)}
            = \exp\!\left(-\tfrac{|\alpha_g|^2+|\alpha_e|^2}{2}+\alpha_g^*\alpha_e\right).
\]
Equation~\eqref{eq:qubitoperator} is a \emph{Kraus map} from $\bxi_Q$
to~$\rho_Q$ with two time-varying Kraus operators:
\[
  M_1(t) = \bP_g + c_{ge}(t)\bP_e, \qquad
  M_2(t) = \sqrt{1-|c_{ge}(t)|^2}\;\bP_e.
\]
One readily checks $M_1^\dag M_1 + M_2^\dag M_2 = \bP_g+\bP_e = \mathbf{I}$,
confirming trace preservation.  Crucially, $\bxi_Q$ evolves according to the
standard SME~\eqref{eq:dynxiQ} with a \emph{constant} efficiency $\eta\in[0,1]$,
thereby avoiding the unphysical detection efficiencies exceeding one that can appear
in the direct non-Markovian description of~$\rho_Q$.

\section{Qudit extension} \label{sec:qudit}

\subsection{Model}

This extension follows~\cite{CrigerEPJQT2016}.  The qubit is replaced by a
finite-dimensional system~$\mathcal{S}$ with orthonormal basis
$\{\ket{s}\}_{s=1,\ldots,n_\mathcal{S}}$.  The SME for the joint density operator is
{\small
\begin{multline*}
  d\rho = -i\!\left[\sqrt{\kappa}\,u\,\ba^\dag+\sqrt{\kappa}\,u^*\ba,\;\rho\right]dt \\
         -i\!\left[\Bigl(\sum_s\chi_s\bP_s\Bigr)\ba^\dag\ba,\;\rho\right]dt
         + \kappa\,\cD_{\ba}(\rho)\,dt
         + \sqrt{\eta\kappa}\,\cM_{\ba}(\rho)\,dw,
\end{multline*}
}
where $\bP_s=\ketbra{s}{s}$ and the $\chi_s$ are real coupling constants.
Each level~$s$ acquires its own coherent pointer amplitude $\alpha_s(t)\in\CC$,
satisfying
\[
  \dot\alpha_s = -i\chi_s\alpha_s - i\sqrt{\kappa}\,u
                 - \tfrac{\kappa}{2}\alpha_s.
\]
The polaron transformation
$\rho = \bigl(\sum_s\bP_s\bD_{\alpha_s}\bigr)\bxi
        \bigl(\sum_s\bP_s\bD_{-\alpha_s}\bigr)$
maps~\eqref{eq:dynrho} (with the qudit dispersive coupling) into
{\small
\begin{multline}\label{eq:dynExt1}
  d\bxi =
  -i\sum_s\left[\bP_s\!\left(\tfrac{i\kappa}{2}
                (\alpha_s\ba^\dag-\alpha_s^*\ba)\right),\bxi\right]dt \\
  -i\sum_s\left[\bP_s\!\left(\Re(\alpha_s\sqrt{\kappa}\,u^*)
                +\chi_s\ba^\dag\ba\right),\bxi\right]dt \\
  + \kappa\,\cD_{\ba+\sum_s\alpha_s\bP_s}(\bxi)\,dt
  + \sqrt{\eta\kappa}\,\cM_{\ba+\sum_s\alpha_s\bP_s}(\bxi)\,dw,
\end{multline}
}
with output
$dy = \sqrt{\eta\kappa}\,\Tr{\bigl(\ba+\sum_s\alpha_s\bP_s\bigr)\bxi
      +\mathrm{h.c.}}\,dt + dw$.

\subsection{Invariant manifold and convergence}

\begin{lemma}[Invariant manifold]\label{lem:ext1}
  If $\bxi(0)=\bxi_\mathcal{S}^0\otimes\ketbra{0}{0}$, then for all $t>0$,
  $\bxi(t)=\bxi_\mathcal{S}(t)\otimes\ketbra{0}{0}$, where $\bxi_\mathcal{S}$
  satisfies
  \begin{multline*}
    d\bxi_\mathcal{S} =
    -i\!\left[\sum_s\Re(\alpha_s\sqrt{\kappa}\,u^*)\bP_s,\;\bxi_\mathcal{S}\right]dt \\
    + \kappa\,\cD_{\sum_s\alpha_s\bP_s}(\bxi_\mathcal{S})\,dt
    + \sqrt{\eta\kappa}\,\cM_{\sum_s\alpha_s\bP_s}(\bxi_\mathcal{S})\,dw,
  \end{multline*}
  with output
  $dy = \sqrt{\eta\kappa}\,\Tr{\bigl(\sum_s\alpha_s\bP_s\bigr)
        \bxi_\mathcal{S}+\mathrm{h.c.}}\,dt + dw$
  and $\bxi_\mathcal{S}(0)=\bxi_\mathcal{S}^0$.
\end{lemma}

\begin{lemma}[Exponential convergence]\label{lem:ext1convergence}
  Every solution of~\eqref{eq:dynExt1} converges exponentially and almost
  surely to $\mathcal{I}_\mathcal{S}=\{\bxi_\mathcal{S}\otimes\ketbra{0}{0}\}$.
  The mean photon number satisfies
  \[
    \frac{d}{dt}\EE{\Tr{\ba^\dag\ba\,\bxi}} = -\kappa\,\EE{\Tr{\ba^\dag\ba\,\bxi}}.
  \]
\end{lemma}

\subsection{State-space input/output representation}

Starting from
$\rho(0)=\rho_\mathcal{S}^0\otimes\ketbra{\alpha^0}{\alpha^0}$,
the state is encoded in $(\{\alpha_s\},\bxi_\mathcal{S})$
with $\alpha_s(0)=\alpha^0$ for all~$s$.
Tracing over $\mathcal{S}$ gives the cavity state
\[
  \rho_C(t) = \sum_s\Tr{\bP_s\bxi_\mathcal{S}(t)}\ketbra{\alpha_s(t)}{\alpha_s(t)}.
\]
Tracing over the cavity gives
\[
  \rho_\mathcal{S}(t) = \sum_{s,s'} c_{s's}(t)\,\bP_s\,\bxi_\mathcal{S}(t)\,\bP_{s'},
\]
where $c_{s's}=\braket{\alpha_{s'}}{\alpha_s}$ is the overlap between
pointer states~$s'$ and~$s$.  The matrix
$C=(c_{s's})_{1\le s',s\le n_\mathcal{S}}$ is the Gram matrix of the
$n_\mathcal{S}$ coherent states $\ket{\alpha_s}$; it is Hermitian positive
semidefinite, so it admits a Hermitian square root
$\sqrt{C}=(r_{s's})$.  Writing $\bK_s(t)=\sum_{s'}r_{s,s'}(t)\bP_{s'}$,
we recover a standard Kraus decomposition
\[
  \rho_\mathcal{S}(t) = \sum_s\bK_s(t)\,\bxi_\mathcal{S}(t)\,\bK_s^\dag(t),
\]
with $\sum_s\bK_s^\dag\bK_s = \mathbf{I}$ (since $\sum_{s'}|r_{s,s'}|^2=1$
for each~$s$).

\section{Multi-cavity extension} \label{sec:multimode}

\subsection{Model}

This section extends the analysis to $n_C$ cavity modes, following
the setup of~\cite{CrigerEPJQT2016,Yu2023StochasticMO}.  Let $\ba_c$
($c=1,\ldots,n_C$) be the annihilation operators of the modes, and let
$\mathcal{S}$ have basis $\{\ket{s}\}_{s=1,\ldots,n_\mathcal{S}}$ as before.
With collective coherent drive~$u$, detunings~$\Delta_c$, dispersive couplings
$\chi_{s,c}$, and collective homodyne detection
$\ba_\mathrm{out}=\sum_c\sqrt{\kappa_c}\,\ba_c$, the SME is
{\small
\begin{multline*}
  d\rho =
  -i\!\left[\sum_c\bigl(\sqrt{\kappa_c}\,u\,\ba_c^\dag
             +\sqrt{\kappa_c}\,u^*\ba_c\bigr),\rho\right]dt \\
  -i\!\left[\sum_c\Bigl(\Delta_c+\sum_s\chi_{s,c}\bP_s\Bigr)
             \ba_c^\dag\ba_c,\rho\right]dt \\
  + \cD_{\sum_c\sqrt{\kappa_c}\ba_c}(\rho)\,dt
  + \sqrt{\eta}\,\cM_{\sum_c\sqrt{\kappa_c}\ba_c}(\rho)\,dw.
\end{multline*}
}
The two-index pointer amplitudes $\alpha_{s,c}(t)\in\CC$ satisfy the coupled ODEs
\begin{equation}\label{eq:alphasc}
  \dot\alpha_{s,c} = -i(\Delta_c+\chi_{s,c})\alpha_{s,c}
                     - i\sqrt{\kappa_c}\,u
                     - \sum_{c'}\frac{\sqrt{\kappa_c\kappa_{c'}}}{2}\alpha_{s,c'}.
\end{equation}
We introduce the polaron transformation with the multi-mode displacement
$\bD_{\alpha_s}=\exp\!\bigl(\sum_c\alpha_{s,c}\ba_c^\dag-\alpha_{s,c}^*\ba_c\bigr)$,
and define the weighted average amplitude
\[
  \sqrt{\bar\kappa}\,\bar\alpha_s = \sum_c\sqrt{\kappa_c}\,\alpha_{s,c},
  \qquad
  \sqrt{\bar\kappa} = \sum_c\sqrt{\kappa_c}.
\]
The transformed equation is
{\small
\begin{multline}\label{eq:dynExt2}
  d\bxi =
  -i\!\left[\sum_s\bP_s\Re(\sqrt{\bar\kappa}\,\bar\alpha_s u^*),\bxi\right]dt \\
  -i\!\left[\sum_s\bP_s\sum_c(\Delta_c+\chi_{s,c})\ba_c^\dag\ba_c,\bxi\right]dt \\
  -i\!\left[\sum_s\bP_s\sum_c
            \frac{i\sqrt{\bar\kappa\kappa_c}}{2}
            (\bar\alpha_s\ba_c^\dag-\bar\alpha_s^*\ba_c),\bxi\right]dt \\
  + \cD_{\sum_s\bP_s(\sqrt{\bar\kappa}\bar\alpha_s+\sum_c\sqrt{\kappa_c}\ba_c)}
    (\bxi)\,dt \\
  + \sqrt{\eta}\,
    \cM_{\sum_s\bP_s(\sqrt{\bar\kappa}\bar\alpha_s+\sum_c\sqrt{\kappa_c}\ba_c)}
    (\bxi)\,dw,
\end{multline}
}
with output
{\small
\[
  dy = \sqrt{\eta}\,\Tr{\sum_s\bP_s\Bigl(\sqrt{\bar\kappa}\bar\alpha_s
       +\sum_c\sqrt{\kappa_c}\ba_c\Bigr)\bxi+\mathrm{h.c.}}\,dt + dw.
\]
}

\subsection{Invariant manifold and convergence}

\begin{lemma}[Invariant manifold]\label{lem:ext2}
  If $\bxi(0)=\bxi_\mathcal{S}^0\otimes\bigl(\otimes_c\ketbra{0}{0}\bigr)$,
  then for all $t>0$,
  $\bxi(t)=\bxi_\mathcal{S}(t)\otimes\bigl(\otimes_c\ketbra{0}{0}\bigr)$,
  where $\bxi_\mathcal{S}$ satisfies
  \begin{multline*}
    d\bxi_\mathcal{S} =
    -i\!\left[\sum_s\Re(\sqrt{\bar\kappa}\,\bar\alpha_s u^*)\bP_s,
              \bxi_\mathcal{S}\right]dt \\
    + \bar\kappa\,\cD_{\sum_s\bar\alpha_s\bP_s}(\bxi_\mathcal{S})\,dt
    + \sqrt{\eta\bar\kappa}\,\cM_{\sum_s\bar\alpha_s\bP_s}(\bxi_\mathcal{S})\,dw,
  \end{multline*}
  with output
  $dy=\sqrt{\eta\bar\kappa}\,\Tr{\bigl(\sum_s\bar\alpha_s\bP_s\bigr)
      \bxi_\mathcal{S}+\mathrm{h.c.}}\,dt+dw$
  and $\bxi_\mathcal{S}(0)=\bxi_\mathcal{S}^0$.
\end{lemma}

\begin{lemma}[Convergence to the invariant manifold]\label{lem:ext2convergence}
  Every solution of~\eqref{eq:dynExt2} tends to converge almost
  surely to $\mathcal{I}_\mathcal{S}$.
  More precisely, the total mean photon number satisfies
  \begin{multline*}
    \frac{d}{dt}\sum_c\EE{\Tr{\ba_c^\dag\ba_c\,\bxi}} \\
    = -\EE{\Tr{\Bigl(\sum_c\sqrt{\kappa_c}\,\ba_c\Bigr)^\dag
               \Bigl(\sum_c\sqrt{\kappa_c}\,\ba_c\Bigr)\bxi}}
    \le 0.
  \end{multline*}
\end{lemma}

\begin{remark}
Unlike the single-mode case, $\sum_c\Tr{\ba_c^\dag\ba_c\,\bxi}$ is not a
strict Lyapunov function for $\mathcal{I}_\mathcal{S}$.  We conjecture that
exponential convergence is equivalent to the requirement that, for $u\equiv 0$
and each~$s$, the linear system~\eqref{eq:alphasc} converges exponentially
to~$0$.  Characterizing the asymptotic regime of~\eqref{eq:dynExt2} from
the spectral properties of~\eqref{eq:alphasc} is an interesting open problem.
\end{remark}

\subsection{State-space input/output representation}

Starting from
$\rho(0)=\rho_\mathcal{S}^0\otimes\bigl(\otimes_c\ketbra{\alpha_c^0}{\alpha_c^0}\bigr)$,
the state is encoded in $\bxi_\mathcal{S}$ (with initial condition
$\rho_\mathcal{S}^0$) and $\alpha_{s,c}$ (with initial condition $\alpha_c^0$,
independent of~$s$).  Tracing over $\mathcal{S}$ gives
\[
  \rho_C(t) = \sum_s\Tr{\bP_s\bxi_\mathcal{S}(t)}
              \Bigl(\otimes_c\ketbra{\alpha_{s,c}(t)}{\alpha_{s,c}(t)}\Bigr).
\]
Tracing over all cavity modes gives
\[
  \rho_\mathcal{S}(t) = \sum_{s,s'} c_{s's}(t)\,\bP_s\,\bxi_\mathcal{S}(t)\,\bP_{s'},
\]
where now $c_{s's}=\prod_c\braket{\alpha_{s',c}}{\alpha_{s,c}}$.  Defining
$C=(c_{s's})$ and its square root $\sqrt{C}=(r_{s's})$ as before, we obtain
the Kraus decomposition
\begin{gather*}
  \rho_\mathcal{S}(t) = \sum_s\bK_s(t)\,\bxi_\mathcal{S}(t)\,\bK_s^\dag(t), \\
  \bK_s(t) = \sum_{s'} r_{s,s'}(t)\,\bP_{s'}.
\end{gather*}

\section{Concluding remarks}

The Markovian state-space formulation developed here---combining a fictitious-qudit
SME, linear ODEs for the pointer amplitudes, and a time-varying Kraus
output map---avoids the unphysical features (negative dephasing rates, efficiencies
exceeding one) of earlier non-Markovian descriptions, while remaining exact on the
invariant manifold.

An important open extension is to account for a finite qubit lifetime~$T_1$:
\begin{multline*}
  d\rho = -i\!\left[\sqrt{\kappa}\,u\,\ba^\dag+\sqrt{\kappa}\,u^*\ba
           +\chi\bSz\ba^\dag\ba,\rho\right]dt \\
         + \left(\kappa\,\cD_{\ba}(\rho)
           + \tfrac{1}{T_1}\cD_{\ketbra{g}{e}}(\rho)\right)dt
         + \sqrt{\eta\kappa}\,\cM_{\ba}(\rho)\,dw.
\end{multline*}
In the regime $\kappa\gg 1/T_1$, a perturbative treatment may yield an
approximate reduced model of the same structure: a fictitious-qubit SME, linear
ODEs, and Kraus output maps.

\subsection*{Acknowledgments}

This work grew out of discussions during the winter of 2019--2020 with
David~DiVincenzo, Benjamin~Huard, Mazyar~Mirrahimi, and Alain~Sarlette.
This project has received funding from the European Research Council~(ERC)
under the European Union's Horizon~2020 research and innovation programme
(grant agreement No.~884762).
This revised version was prepared with the assistance of Claude
(Anthropic), which contributed to improving the English, enriching the
bibliography, verifying the formulas, and enhancing the overall pedagogical
presentation.

\end{document}